 \documentclass[a4paper,ngerman,UKenglish,12pt]{scrartcl}
 
 \usepackage[margin=1in]{geometry}
 \usepackage[utf8]{inputenc}
 \usepackage{braket}
 \let\originalleft\left
 \let\originalright\right
 \renewcommand{\left}{\mathopen{}\mathclose\bgroup\originalleft}
 \renewcommand{\right}{\aftergroup\egroup\originalright}
 \usepackage{breakurl}
 \usepackage{amsmath}
 \usepackage{amsfonts}
 \usepackage{dsfont}
 \usepackage{hyperref}
 \usepackage{tikz}
 \usepackage{lmodern}
 \usepackage{lettrine}
 \input Zallman.fd

 \usepackage[newzealand]{babel}
 \usepackage{microtype}
 \usepackage[T1]{fontenc}
 \usepackage{csquotes}
 \usepackage[sorting=anyt,doi=true,eprint=true,arxiv=abs,hyperref=true,style=nature-modified,citestyle=alphabetic]{biblatex}
 \DeclareNameAlias{default}{given-family}
 \defbibenvironment{bibliography}
 {
 	\list
 	{\printtext[labelalphawidth]{%
 			\printfield{labelprefix}%
 			\printfield{labelalpha}%
 			\printfield{extraalpha}}}
 	{\setlength{\labelwidth}{\labelalphawidth}%
 		\setlength{\leftmargin}{\labelwidth}%
 		\setlength{\labelsep}{\biblabelsep}%
 		\addtolength{\leftmargin}{\labelsep}%
 		\setlength{\itemsep}{\bibitemsep}%
 		\setlength{\parsep}{\bibparsep}}%
 	}
 {\endlist}
 {\item}
 \definecolor{lime}{HTML}{A6CE39}
 \definecolor{cinnabar}{rgb}{0.89, 0.26, 0.2}
 
 \newcommand{\orcidicon}{%
 	\begin{tikzpicture}
 	\draw[lime, fill=lime] (0,0) 
 	circle [radius=0.16] 
 	node[white] {{\fontfamily{qag}\selectfont \tiny ID}};
 	\draw[white, fill=white] (-0.0625,0.095) 
 	circle [radius=0.007];
 	\end{tikzpicture}
 	\hspace{-3mm}
 }
 \newcommand\orcidSeSc{{\href{https://orcid.org/0000-0003-1997-0026}{\orcidicon}}}
 
 \setcapindent{0em}
 \allowdisplaybreaks
 
 \newcommand{\kl}[1]{\left( #1 \right)}
 \newcommand{\kle}[1]{\left[ #1 \right]}
 \newcommand{\ed}[1]{\frac{1}{#1}}
 
 \newcommand{\defi}{\mathrel{\mathop:}=}
 \newcommand{\ifed}{=\mathrel{\mathop:}}
 \newcommand{\dif}{\ensuremath{\operatorname{d}}\!}
 
 \title{Frenemies with Physicality: Manufacturing Manifold Metrics}
 \author{{\small}{Sebastian Schuster*\orcidSeSc}\\\small (*): \url{sebastian.schuster@utf.mff.cuni.cz}\\\small Institute of Theoretical Physics,
 			Faculty of Mathematics and Physics, \\\small
 			Charles University,\\\small
 			V~Hole\v{s}ovi\v{c}k\'{a}ch~747/2,
 			180 00 Prague 8,
 			Czech Republic.}
 \date{\small Submission date: \textbf{29 March 2023}\\ Essay written for the Gravity Research Foundation 2023 Awards for Essays on Gravitation.}
 
\begin{document}
 	\maketitle
 	\begin{abstract}
 		Physicality has the bad habit of sneaking up on unsuspecting physicists. Unfortunately, it comes in multitudinous incarnations, which will not always make sense in a given situation. Breaching a warp drive metric with physical arguments is all good, but often what counts as physicality here is but a mere mask for something else. In times of analogue space-times and quantum effects, a more open mind is needed. Not only to avoid using a concept of physicality out of its natural habitat, but also to find useful toy models for our enlarged phenomenology of physics with metrics. This journey is bound to be as vexing, confusing, and subtle as it (hopefully) will be illuminating, entertaining, and thought-provoking.
 	\end{abstract}
 	\clearpage
 	
 	\section{To Integrate or to Differentiate?}
 	\lettrine[lines=4, findent=3pt, nindent=0pt]{T}{here is the old adage} that differentiation is a skill, while integration is an art. This saying is fairly general, but certainly also true in the context of the Einstein equations of general relativity. To view solving these equations as a matter of integrating a non-linear system of partial differential equations is just as traditional as the opening adage. However, complicated questions need to be settled first, for example, which formulation is appropriate in a given setting, or how to find rigorously the two propagating degrees of freedom in the ten degrees of freedom of the metric itself.
 	
 	However\dots One could just pick a metric, differentiate twice, sum appropriately, and end up with a stress-energy tensor $T_{\mu\nu}\defi 8\pi G/c^4 G_{\mu\nu}$ that trivially satisfies the Einstein equation. Put bluntly, any metric is a solution of \emph{some} Einstein equation. Let us call this \enquote{reverse engineering}---though it might tickle the fancy of sci-fi aficionados a tad too much.
 	
 	This certainly seems like a preposterous approach. The goal of this essay is to challenge this appearance. Quite the contrary, I will argue \emph{that} skill may very well be as artistic as the integration that purports to be art from the get-go. Still, objections can be found aplenty, and are particularly encountered in extemporaneous, more or less disparaging verbal communication in the wild: Used are various combinations of classical theorems, energy conditions, or just plain disgust at the proliferous chaff of additional metrics this lets in. Weirdly enough, as many objections are not as vigorously put forward in our other, older, classical field theory: Homogeneous magnetic fields are as unphysical as they are widely and \emph{experimentally successfully} applied\dots
 	
 	Pursuing this goal (or, perhaps, pet peeve?), one can hardly avoid invoking energy conditions, particularly those of the pointwise persuasion. To spoil the joy of discovering the conclusion in due course: They have their uses---but especially when applied to the usual exemplars of reverse-engineered metrics, they can turn interesting issues of physicality into flashy, superficial, often faulty equation juggling for academic martinets.
 	
 	Lastly, I would like to add an unsurprising disclaimer: Admittedly, I count myself among the above-denigrated science fiction aficionados. Yet as much as I love the genre's tropes and questions, and as much as its technical tropes can provide fascinating physics questions---this application is usually a case of \emph{breaking} things, and trying to salvage as much as possible from the wreckage\dots And while toddlers and particle physicists alike are united in vouching for destructive processes as a pathway to learning, problems creep in when the deconstruction is blindly interpreted as constructive, or the theoretical as practical.
 	
 	With this short introduction out of the way, let me guide the willing reader through the remainder: First, I will go into a bit more detail about the status of energy conditions in section~\ref{sec:disabuse}. Next, I will belabour my frustrations of people linking energy conditions monolithically with the physicality of a metric some more with an explicit example in section~\ref{sec:drain}. In section~\ref{sec:beyond}, I endeavour to show that this example has relevance beyond its decidedly un-gravitational origin, concerning all toy models with Lorentzian geometry. Finally, I will deliver the ambiguous catastrophe for our anti-hero, Physicality, in section~\ref{sec:theend}.
 	
 	\section{To Disabuse of Energy Conditions}\label{sec:disabuse}
 	Energy conditions are great, and amazing, and important \cite{1405.0403,2003.01815}. In particular, they allow to derive many classical theorems of \emph{mathematical} general relativity: Singularity theorems, area theorems, positive mass theorems, geometric and analytic properties of apparent horizons, topological censorship, \dots \cite{Hawking:1973uf,ClassQuantGrav.12.L99,0811.4721,Choquet-Bruhat:2009xil}
 	
 	Now that I have said this, I can focus on the small print. Many of these \emph{mathematical} questions have to do with what is physical or accessible through physical means---within the framework of general relativity. A statement guaranteeing that mass stays positive certainly can be seen as confirming our expectations of what a \enquote{physical} situation should be. In this light it is not too surprising that the early history of energy conditions had to do with how to avoid invoking relativistically ill-behaved equations of state in general relativity \cite{SovPhysJETP.14.1143}. At least two of the above-mentioned theorems, though, seem to go in a rather awkward direction if we wanted energy conditions to validate our dearly-held beliefs: The singularity theorems are \emph{the} point when quantum gravity gets invoked to save us all from the evils of these predicted, pesky singularities. Likewise, our inability to probe the topology of space-time employing only matter satisfying energy conditions is rather vexing (and non-quantum \cite{1706.08978}). Then again, existence and uniqueness results concerning partial differential equations governing marginally outer trapped (open) surfaces don't really point our physical intuition one way or the other; the concepts encountered in this context are either too intrinsic to the mathematical structures of general relativity, or better addressed by theorems aiming straight for our intuition.
 	
 	This is not to mean that the theorems fail! Rather, it is to point out that physicality is often a stand-in for beliefs about what physics should do based on other theories' results. Take earlier encounters with singularities: They correctly pointed out when things go wrong. This was addressed both by new theories superseding the Old Ones, as well as by developing mathematical tools to brave the onslaught of a singularity. (To add a bit of jargon and name-dropping, take Schwartz' distribution theory or Colombeau's algebras.) In particular, these theorems collectively point to the fact that different notions of what we deem physical can clash: Do we want positive mass or singularities?
 	
 	In a very similar vein---mathematically, philosophically, and physically---physical intuition in general relativity can clash with itself when the notion of \enquote{inextendibility} meets other \enquote{physical} requirements (global hyperbolicity, geodesic completeness, \dots) \cite{Manchak2021CollectingCollections}. Now, obviously, the solution is not to throw one's hands in the air, despair, and quit relativity\footnote{Though it would allow one to fight \emph{even more} against the oncoming climate crisis.}---it will sadly not resolve this very pressing conundrum. Rather, it goes to show what happens when one moves outside the realm of validity of different \enquote{physicalities}. Especially if this realm of validity has not yet been fully established. We can be reasonably certain that general relativity is not valid at Planck scale physics, so the presence of singularities should warn us of its terminology becoming meaningless. Even away from such singularities, though, it bears remembering that many a notion of general relativity comes with the added baggage of assuming things like global hyperbolicity, asymptotic flatness, vacuum, symmetries, \dots The simplest example is how anti-deSitter does not allow a (straightforward) initial value formulation, as it is not globally hyperbolic. Mass is hard to pinpoint if one tries to evade this (or other issues) by simply going local.
 	
 	This is a good point to come back to energy conditions: Point-wise or not, they are closely tied to physical intuition about mass. This is most obvious for the weak energy condition and a Hawking--Ellis type~I stress-energy tensor. \emph{Part} of the resulting expressions involving pressures $p_i$ and energy density $\rho$ is a \enquote{$\rho>0$}. This is blissfully close to \enquote{mass should be positive.} If the given stress-energy tensor is \emph{not} of type~I, this becomes trickier.\footnote{Additionally, this is the appropriate moment to point out that all too often the \enquote{for all four-velocities} (to be interpreted as: \enquote{for all observers}) part of pointwise energy conditions is forgotten when checking their validity for any given metric (reverse-engineered or not.) \cite{2105.03079}}
 	
 	To make matters worse, we have theoretical and \emph{experimental} examples that violate (pointwise) energy conditions---even the weakest of the famous extant ones, the null energy condition \cite{gr-qc/0205066}. And unlike the above, glib example of the \enquote{homogeneous magnetic field}, there is no invocation of things being a \enquote{local approximation} or of gluing techniques to save the day. The energy conditions' range of validity is \emph{pointwise!}
 	
 	So, subsequent developments in energy conditions have aimed at ridding them of their pointwise quality. The resulting, pluriform \emph{averaged} energy conditions succeeded in weakening the conditions of classical theorems of mathematical relativity; perhaps more importantly, they paved the way for quantum energy inequalities (QEIs). QEIs provide us with bounds on how much quantum fields can violate energy conditions. Additionally, this allows to exchange the ingredient of \enquote{and now add matter that \emph{somehow} satisfies these ad-hoc inequalities} in theorems by an assumption of the type \enquote{with \emph{this} kind of quantum matter}. See, for example, the semi-classical singularity theorem \cite{2108.12668}. A possible future application of this would be to (finally) test whether quantum fields themselves allow for the various reverse-engineered solutions violating pointwise energy conditions. Yet in light of the growing number of semi-classical singularity theorems, even for something as \enquote{benign} as regular black holes, it seems unlikely that quantum matter is the much-evoked answer for arbitrarily advanced civilizations to engineer such reverse-engineered metrics. It seems to be just as unlikely for mere curved space-time quantum field theory to protect us from singularities; quantum gravity still seems to rear its spectre-like head in the corners of our calculational sight. With it, the realm of validity of energy conditions\footnote{Be it classical relativity or curved space-time quantum field theory; back-reaction or no back-reaction.} seems not to be the realm for answering \emph{all} the questions of physicality that general relativity puts in front of us.
 	
 	\section{Metrics without Energy Conditions?}\label{sec:drain}
 	Let us take a step back: Energy conditions or related inequalities---classical or quantum---all pertain to the Einstein equations or some generalization thereof. We need some way to relate the left-hand-side involving the metric and its derived geometric tensors to the right-hand-side, the source term, which in general relativity is just the stress-energy tensor. Essentially, this means one has a metric theory (or some other more or less directly related geometric picture for this theory). This then is spiced up with notions of frames, observers, and frame-changes. The spice is why energy conditions are always decorated with at least one $\forall$-quantifier to cover all possible observers' perspectives.
 	
 	Not all metrics in physics, however, have an origin of this type. Take analogue space-times. Most analogue space-times (at least all I can think of) have a preferred frame and no simple notion of frame-independence as general relativity (and company) does (and do). Of course, one could weaken the interpretation of covariance to a simple statement about re-expressing given equations in new coordinates---quite generally possible---but this would not introduce new observers, just (complicated) cartography for the original observer describing the analogue metric system.
 	
 	In absence of all these observers, the question of physicality of a given metric cannot be hoisted upon energy conditions---and following the previous section, even \emph{with} an observers at hand, their violation is at best a warning sign of uncanny physics. Convergence conditions, equivalent to energy conditions in Einstein's theory of gravity, become even flimsier, as the link between geometry and matter becomes more tenuous.
 	
 	As a proof of principle of this observation, let us take a look at the poster child of fluid analogues---the draining bathtub metric.\footnote{Also known, particularly within fluid dynamics, under the (at first glance and etymologically endearingly oxymoronic) name \enquote{irrotational vortex.}} Here, the fluid flow \cite{gr-qc/0505065}
 		\begin{equation}
 			\mathbf{v} = -\nabla \phi_0 = \frac{A \hat{r} + B \hat{\theta}}{r},\label{eq:irrvortflow}
 		\end{equation}
 	where $A, B$ are parameters corresponding to radial flow and tangential flow, respectively, and $\phi_0$ is its velocity potential, will induce an effective, analogue metric
 		\begin{align}
 			\dif s^2 &= -\frac{\rho}{c_{\text{s}}}\kle{\kl{c_{\text{s}}^2-\frac{A^2+B^2}{r^2}}\dif t^2 - 2\frac{A}{r} \dif r \dif t - 2B \dif \theta \dif t + \dif r^2 + r^2 \dif \theta^2 + \dif z^2}.\label{eq:irrvortmetric}\\
 			&\ifed \hphantom{-}g^{\text{eff}}_{\mu\nu} \dif x^\mu \dif x^\nu \nonumber
 		\end{align}
 	Here, $\rho$ is the fluid's density, and $c_{\text{s}}$ is its speed of sound. More concretely, this metric appears since the potential flow of sound waves, in the sense of being perturbations $\phi_1$ on top of the flow of an irrotational vortex \eqref{eq:irrvortflow}, will have to follow the Laplace--Beltrami equation
	 	\begin{equation}
	 		\Box \phi_1 \defi \ed{\sqrt{-g_{\text{eff}}}}\partial_\mu \kl{\sqrt{-g_{\text{eff}}} \, g^{\mu\nu}_{\text{eff}} \partial_\nu \phi_1} =0.\label{eq:LBeqn}
	 	\end{equation}
 
 	Now what would happen if one were to look at the weak energy condition
 		\begin{equation}
 			\forall \text{~four-velocities~} W: \quad T_{\mu\nu} W^\mu W^\nu \geq 0?
 		\end{equation}
 	In the derivation of the Laplace--Beltrami equation~\eqref{eq:LBeqn}, the Euler equation entered, so we are dealing with an ideal fluid. There is the equation-of-state to play with, as this is not specified at this stage (and influences, for example, the speed of sound $c_{\text{s}}$). So, in general one could not say much, and would just refer to the usual folk wisdom that \emph{obviously} realistic, \emph{ideal} fluids \emph{in a lab} would satisfy the weak energy condition.
 	
 	However, one could look at it from a different point of view. If this draining bathtub metric were to occur in a general-relativistic setting, what would the stress-energy tensor have to be? Obviously, this would be rather counter-factual, as the metric did not arise from solving the Einstein equation, but let's do it anyway. Calculating the \enquote{effective} Einstein tensor $G^{\text{eff}}_{\mu\nu}$ proportional to the \emph{effective} stress-energy tensor is straightforward. The fluid flow comes with a corresponding four-velocity
	 	\begin{equation}
	 		V^\mu \defi \ed{\sqrt{\rho c_{\text{s}}}} \begin{pmatrix}
	 			1\\
	 			\mathbf{v}
	 		\end{pmatrix},
	 	\end{equation}
 	which fulfils the usual $ V^\mu V^\nu g_{\mu\nu}^{\text{eff}} = -1$. Contracting this with the Einstein tensor gives
 		\begin{equation}
 			G^{\text{eff}}_{\mu\nu} V^\mu V^\nu = -\frac{A^2+B^2}{r^4 \rho c_{\text{s}}}.
 		\end{equation}
 	Already the four-velocity of the fluid flow itself is apparently enough to violate the energy condition. Everywhere. Still, no one would consider this metric to be less valuable for it; this stress-energy tensor is only ever encountered if one were to consider the metric as arising from general relativity. It has a different origin, will have a different stress-energy tensor, and has its place and value apart from general relativity proper. Its Lorentzian geometry still has physics to teach, though: It serves as a good example of horizons in non-gravitational settings.
 	
 	\section{Beyond Mere Analogues}\label{sec:beyond}
 	Such non-gravitational, metric encounters still have and should have an imprint of our understanding of \enquote{actual} gravitational metrics. While there is an ever-growing plethora of metrics already to be found in gravity---especially once one allows for modified theories---they can be subdivided into three broad, sometimes overlapping classes: The Standard~[sic!], the special interest, and the mind-numbing. The Standard are metrics found in a hypothetical, global average of general relativity courses (Schwarzschild, Kerr, (anti-)deSitter, Friedmann--Lemaître--Robertson--Walker, \dots[?]). Of \enquote{special interest} are those that are (according to some) still analytically tractable and serve a particular purpose. Examples abound: $C$-metric, Lemaître--Tolman--Bondi cosmologies, Morris--Thorne wormholes, Gott and Gödel, and many, many more. The \enquote{mind-numbing} would include all sorts of approximate or numerical metrics, like Brill--Lindquist initial data: A standard, numerical toy model at the intersection of mind-numbing and special interest.
 	
 	This mind-numbing group will require a large class of toy models to study for specific questions: The absence of (manageable) analytical results means that our intuition quickly takes flight to greener pastures. Yet, questions of trapped surfaces, \enquote{shadows} (a.k.a. silhouettes), orbits, and many more things of physical relevance in gravitational astrophysics need \emph{some} intuition. Reverse-engineered toy models, even those inspired by other fields of physics (as by analogues), can aide this greatly. Even and especially if they violate dearly-held concepts of \enquote{physicality.} Likewise, analogues themselves will benefit from a larger class of toy models to draw from. Concepts like \enquote{rainbow metrics} and their close connection to more general dispersion relations indicate that, at the end of the day\footnote{Or however long the analysis took\dots}, we may encounter concepts not found in standard general relativity. \emph{Vice versa,} these concepts can come to our aid when general relativity uses its non-linearity to trip us up and trap us. Or when its non-linearity is insufficient and we need to deal with modified theories. But while mathematicians certainly seem to enjoy the freedom of their Platonic world when considering compact Lorentzian manifolds as nice objects to study---yet many physicists seem to be more hesitant to come up with new Lorentzian metrics.
 	
 	This is not to say that such toy models don't exist to some extent already. Still, the various toy models carry \emph{almost all} additional baggage of physicality inspired by gravity. Take collapse and evaporation: The Vaidya metric provides a simple metric to study this, reducing the problem to the Schwarzschild case plus a null shell. Collisional toy models, like the above mentioned Brill--Lindquist initial data, still have to rely on the numerical solution of the Einstein equations, even when the question is concerned (primarily) with geometric features like apparent horizons or marginally trapped surfaces \cite{2104.10265}.
 	
 	I would like to advocate for a more enthusiastic employment of reverse-engineering: By constructing metrics with specific properties in mind, irrespective of field equations, it is possible to study many of the questions encountered in both analogue and gravitational space-times in a more model-independent way. I am aware of the danger of even more metric proliferation\footnote{As scholars prefer eponyms to more telling naming schemes, any proliferation of this kind is likely to further remove science from being human-readable or -learnable. \enquote{Falling raindrop coordinates} is, in my humble opinion, vastly preferable to \enquote{Painlevé--Gullstrand coordinates}. Alas, this and many similar battles seem to have been mostly lost\dots \cite[\S19]{Halmos1973HowTo}}, and to some extent I share people's concern. But someone's \enquote{metric too much} might still end up being someone else's \enquote{invaluable inspiration.}
 	
 	To give a small example of the difficulty in deciding a metric's relevance, let me attempt an \emph{ad hoc} construction of a toy model. In super-luminal warp drives, one can encounter non-compact horizons (with links to various entertaining problems of physicality). Given the free functions in their various incarnations and the rather ugly algebra in most of the explicit examples, studying this null surface is non-trivial. However, taking inspiration from the Schwarzschild metric, the metric
 		\begin{equation}
 			\dif s^2 = -f(x)\dif t^2 + f(x)^{-1}\dif x^2 + \dif y^2 + \dif z^2 \label{eq:generalplanarhorizon}
 		\end{equation}
 	easily can provide us with a non-compact, flat horizon to study, if at some $x_{\text{h}}$, we have $f(x_{\text{h}})=0$. Depending on one's needs, we can have really uncomfortable singular surfaces ($f(x) \defi 1-x_{\text{h}}/x$), or something more regular ($f(x) \defi [a \exp(-ax)-1$],\footnote{With the Lambert $W$-function, the parameter $a$ is related to the horizon location of the (single) horizon surface by $W(-x_{\text{h}})/(-x_{\text{h}})$. A strategically added square makes things even more regular, but yields in general two horizons.} \cite{GenRelGrav.24.433}).
 	
 	Without the physical principles that allowed us to derive the Schwarzschild metric, the general form~\eqref{eq:generalplanarhorizon} is the best one can do. Further fixing of the toy model (as in the previous paragraph) will have to mind the task at hand. What is its context? Which calculations and comparisons become particularly simple for what $f(x)$? Ideally, comparisons to the (here unknown) problem at hand would leave $f(x)$ free for as long as possible. Yet even this toy model is already much more accessible to experience and intuition. However, before I overstay the reader's welcome---let's wrap up.
 	
 	\section{A Subtle Razor}\label{sec:theend}
 	With the advent of analogue space-times, metrics have been rapidly propelled beyond their cradle in general relativity. Meanwhile, in astrophysics, gravitational wave astronomy made numerical relativity ever more important. Yet our understanding of how \enquote{physicality} constraints a metric has moved little, despite philosophers pointing out that various notions of physicality are mutually incompatible, and a tacit, unspoken agreement among physicists that different circumstances require different \enquote{physicalities.}
 	
 	I firmly believe (and try to correct with this essay) that the implicit nature of this agreement holds relativity back---in astrophysics and elsewhere. Admittedly, the flamboyant noise many engineered metrics of science-fiction fame make when confronted with these traditional notions of physicality have been, are, and will be distracting. Nevertheless, toy models that simply help us understand a particular phenomenological aspect of \emph{some} feature of \emph{some} space-time will greatly benefit from not being constrained by physicality, when they were not meant to study a physical situation, just a specific effect.
 	
 	This is nothing new. Looking at classical field theories, one has the example of the homogeneous magnetic field. In quantum mechanics, much of the mathematics physicists are exposed to is the language of bounded operators on a Hilbert space, when few physical models come without an unbounded operator. So, much of our intuition about the quantum world is informed by an often incomplete understanding of the mathematical foundation. Yet many toy models, like those of a qubit, \emph{are} bounded. Then Hilbert space is even finite-dimensional. Quantum gravity, at least the canonical version, goes the extra Planck length to the other end of the spectra [sic!]: The Hamiltonian constraint of diffeomorphism-invariant theories forces us to look into the void of $H\ket{\psi}=0$, and ask hard interpretational questions. Cooking up mini-superspace toy models is par for the course. Toy models study physicality itself in absence of an established notion of it.
 	
 	Physicality has its uses. Even the very flimsy notion provided through energy conditions serves as a warning sign under the right circumstances. Yet we should be wary of physicality, lest it confines our imagination, our intuition---or the lessons and fun of breaking physics: \enquote{True humour begins when one ceases to take one's own [field] seriously.} \cite{HesseSteppenwolf}
 	
 	\appendix
 	\section*{Acknowledgments}
 	Thanks for stimulating discussions, debugging, and/or proof-reading go to Ana Alonso Serrano, Kam To Billy Chan, Aindriú Conroy, Finnian Gray, Sk Jahanur Hoque, David Kofroň, Eleni Kontou, Petr Kotlařík, Pavel Krtouš, David Kubizňák, Carlos Peón-Nieto, Jessica Santiago, Matt Visser, and many more. Apologies to anyone I forgot! The author acknowledges support from the technical and administrative staff at the Charles University, and financial support from Czech Science Foundation grant GACR~23-07457S.
 	
 	\printbibliography

@InCollection{1405.0403,
  author       = {Erik Curiel},
  booktitle    = {{Towards a Theory of Spacetime Theories}},
  publisher    = {Springer},
  title        = {{A Primer on Energy Conditions}},
  year         = {2017},
  chapter      = {3},
  editor       = {Dennis Lehmkuhl and Gregor Schiemann and Erhard Scholz},
  isbn         = {978-1-4939-3209-2},
  pages        = {43--104},
  series       = {Einstein Studies},
  volume       = {13},
  doi          = {10.1007/978-1-4939-3210-8_3},
  eprint       = {1405.0403},
  eprinttype   = {arxiv},
  ids          = {CurielPrimerEnergyConditions},
  primaryclass = {physics.hist-ph},
}

@Article{0811.4721,
  author       = {Lars Andersson and Marc Mars and Jan Metzger and Walter Simon},
  journal      = {Classical and Quantum Gravity},
  title        = {{The time evolution of marginally trapped surfaces}},
  year         = {2009},
  month        = apr,
  number       = {8},
  pages        = {085018},
  volume       = {26},
  doi          = {10.1088/0264-9381/26/8/085018},
  eprint       = {0811.4721},
  eprinttype   = {arxiv},
  ids          = {AnderssonEtAl2009MOTSEvo},
  primaryclass = {gr-qc},
}

@Article{2108.12668,
  author       = {Christopher J. Fewster and Eleni-Alexandra Kontou},
  journal      = {Classical and Quantum Gravity},
  title        = {{A semiclassical singularity theorem}},
  year         = {2022},
  month        = mar,
  number       = {7},
  pages        = {075028},
  volume       = {39},
  doi          = {10.1088/1361-6382/ac566b},
  eprint       = {2108.12668},
  eprinttype   = {arxiv},
  ids          = {FewsterKontou2021SemiClassSingThm},
  primaryclass = {gr-qc},
}

@Article{GenRelGrav.24.433,
  author  = {Robert B. Mann},
  journal = {General Relativity and Gravitation},
  title   = {{Lower Dimensional Black Holes}},
  year    = {1992},
  month   = {4},
  pages   = {433--449},
  volume  = {24},
  doi     = {10.1007/BF00760418},
  ids     = {Mann1992LowDimBH},
}

@Article{ClassQuantGrav.12.L99,
  author  = {Gregory J. Galloway},
  journal = {Classical and Quantum Gravity},
  title   = {On the topology of the domain of outer communication},
  year    = {1995},
  month   = jul,
  number  = {10},
  pages   = {L99--L101},
  volume  = {12},
  doi     = {10.1088/0264-9381/12/10/002},
  ids     = {Galloway1995TopoCensor},
}

@Article{gr-qc/0505065,
  author     = {Carlos Barceló and Stefano Liberati and Matt Visser},
  journal    = {Living Reviews in Relativity},
  title      = {{Analogue Gravity}},
  year       = {2011},
  number     = {3},
  volume     = {14},
  doi        = {10.1007/lrr-2011-3},
  eprint     = {gr-qc/0505065},
  eprinttype = {arxiv},
  ids        = {lrrAnalogue},
}

@Book{HesseSteppenwolf,
  author        = {Hermann Hesse},
  publisher     = {Suhrkamp},
  title         = {{Der Steppenwolf}},
  year          = {2001},
  edition       = {Edition Suhrkamp},
  isbn          = {978-3-518-41690-7},
  note          = {First published by S. Fischer Verlag, 1927.},
  afterword     = {Volker Michels},
  hyphenation   = {german},
  language      = {german},
  origpublisher = {S. Fischer Verlag},
  origyear      = {1927},
  shorthand     = {Hes27},
}

@Article{gr-qc/0205066,
  author       = {Carlos Barcelo and Matt Visser},
  journal      = {International Journal of Modern Physics D},
  title        = {{Twilight for the energy conditions?}},
  year         = {2002},
  month        = mar,
  number       = {10},
  pages        = {1553--1560},
  volume       = {11},
  doi          = {10.1142/S0218271802002888},
  eprint       = {gr-qc/0205066},
  eprinttype   = {arxiv},
  ids          = {Twilight},
  primaryclass = {gr-qc},
}

@Article{2104.10265,
  author       = {Daniel Pook-Kolb and Robie A. Hennigar and Ivan Booth},
  journal      = {Physical Review Letters},
  title        = {{What Happens to Apparent Horizons in a Binary Black Hole Merger}},
  year         = {2021},
  month        = oct,
  number       = {18},
  pages        = {181101},
  volume       = {127},
  doi          = {10.1103/PhysRevLett.127.181101},
  eprint       = {2104.10265},
  eprinttype   = {arxiv},
  ids          = {Pook-KolbEtAl2021AppHorMerger},
  primaryclass = {gr-qc},
}

@Book{Hawking:1973uf,
  author    = {Stephen William Hawking and George Francis Rayner Ellis},
  publisher = {Cambridge University Press},
  title     = {{The large scale structure of space-time}},
  year      = {1974},
  isbn      = {978-0-521-09906-6},
  series    = {{Cambridge Monographs on Mathematical Physics}},
  doi       = {10.1017/CBO9780511524646},
  ids       = {HawEll},
}

@Article{SovPhysJETP.14.1143,
  author  = {Yakov Borisovich Zel'dovich},
  journal = {Soviet Physics JETP},
  title   = {{The equation of state at ultrahigh densities and its relativistic limitations}},
  year    = {1962},
  month   = may,
  note    = {Translation of Zhur. Eksptl'. i Teoret. Fiz. 41 (1961) 1609--1615 by W.H.~Furry},
  number  = {5},
  pages   = {1143--1147},
  volume  = {14},
  ids     = {Zeldovich61TraceECViolation},
  url     = {http://jetp.ras.ru/cgi-bin/e/index/r/41/5/p1609?a=list},
}

@Article{1706.08978,
  author      = {Keith K. Ng and Robert B. Mann and Eduardo Martín-Martínez},
  journal     = {Physical Review D},
  title       = {{Over the horizon: Distinguishing the Schwarzschild spacetime and the $\mathds{RP}^3$ spacetime using an Unruh--DeWitt detector}},
  year        = {2017},
  month       = {10},
  number      = {08},
  pages       = {085004},
  volume      = {96},
  doi         = {10.1103/PhysRevD.96.085004},
  eprint      = {1706.08978},
  eprintclass = {gr-qc},
  eprinttype  = {arxiv},
  ids         = {NgMannMartinMartinez2017ProbingTopo},
}

@Article{2105.03079,
  author       = {Jessica Santiago and Sebastian Schuster and Matt Visser},
  journal      = {Physical Review D},
  title        = {Generic warp drives violate the null energy condition},
  year         = {2022},
  month        = mar,
  number       = {6},
  pages        = {064038},
  volume       = {105},
  doi          = {10.1103/PhysRevD.105.064038},
  eprint       = {2105.03079},
  eprinttype   = {arxiv},
  ids          = {SantiagoSchusterVisser2021NoWarp},
  primaryclass = {gr-qc},
}

@Book{Choquet-Bruhat:2009xil,
  author    = {Yvonne Choquet-Bruhat},
  publisher = {Oxford University Press},
  title     = {{General Relativity and the Einstein Equations}},
  year      = {2009},
  isbn      = {978-0-19-923072-3},
  series    = {{Oxford Mathematical Monographs}},
  ids       = {ChoBru},
}

@InCollection{Halmos1973HowTo,
  author    = {Paul Richard Halmos},
  booktitle = {{How to Write Mathematics}},
  publisher = {American Mathematical Society},
  title     = {{How to Write Mathematics}},
  year      = {1973},
  editor    = {Norman Earl Steenrod and Paul Richard Halmos and Menahem Max Schiffer and Jean Alexandre Eugène Dieudonné},
  isbn      = {0-8218-0055-8},
  pages     = {19--48},
}

@Article{2003.01815,
  author       = {Eleni-Alexandra Kontou and Ko Sanders},
  journal      = {Classical and Quantum Gravity},
  title        = {{Energy conditions in general relativity and quantum field theory}},
  year         = {2020},
  month        = sep,
  number       = {19},
  pages        = {193001},
  volume       = {37},
  doi          = {10.1088/1361-6382/ab8fcf},
  eprint       = {2003.01815},
  eprinttype   = {arxiv},
  ids          = {KontouSanders2020ECReview},
  primaryclass = {gr-qc},
}

@InCollection{Manchak2021CollectingCollections,
  author    = {family=Manchak, given=JB, given-i={JB}},
  booktitle = {{Hajnal Andréka and István Németi on the Unity of Science: From Computing to Relativity Theory Through Algebraic Logic}},
  publisher = {Springer},
  title     = {{General Relativity as a Collection of Collections of Models}},
  year      = {2021},
  chapter   = {19},
  editor    = {Judit Madarász and Gergely Székely},
  isbn      = {978-3-030-64186-3},
  pages     = {409--425},
  series    = {{Outstanding Contributions to Logic}},
  volume    = {19},
  doi       = {10.1007/978-3-030-64187-0_17},
  url       = {http://philsci-archive.pitt.edu/14882/},
}
 \end{document}